\documentclass{du-journals}
\usepackage{amsmath,amssymb,bm}
\usepackage{graphicx}
\usepackage{hyperref}
\usepackage{booktabs}
\usepackage{tikz}
\usetikzlibrary{arrows.meta,decorations.pathmorphing}

\title{Gravitational-Wave Modulation of Atomic Spontaneous Emission:\\
Electrodynamical Theory, Quadrupolar Redistribution, and Rydberg Cavity Metrology}
\subtitle{Gravitational Wave interaction with Atom}

\newcommand{\plainauthorA}{Sohrab Rahvar}

\author{Sohrab Rahvar}
\address{Department of Physics, Sharif University of Technology, P.O. Box 11365-9161,\\
Tehran, Iran\\
Research Center for High Energy physics , Sharif university of Technology , Tehran , Iran \\
rahvar@sharif.edu}

\begin{document}



\begin{abstract}
Studying Gravitational Waves (GWs) provide a unique probe of the interplay between quantum matter and the dynamics of spacetime. We develop a QED framework for spontaneous emission in the presence of a weak classical GW by quantizing the electromagnetic field in a linearized gravitational background. We show that a passing GW dynamically phase-modulates vacuum electromagnetic modes, dressing photon states into a Floquet ladder of sidebands separated by the GW frequency. While the total spontaneous decay rate remains unchanged to first order in the GW strain, the emission pattern develops a characteristic spin-2 quadrupolar angular redistribution modulated with the GW phase. For finite interaction times, interference between the carrier and Floquet sidebands generates phase-dependent spectral asymmetries in the transient emission spectrum. To explore experimental observability, we formulate a cavity-QED master equation description and propose a differential cross-cavity detection protocol employing Rydberg ensembles or highly charged ions. These results establish cavity QED as a promising platform for exploring quantum electrodynamics in dynamical spacetime and for high-frequency gravitational-wave detection.
\end{abstract}



\section{Introduction}

The direct detection of gravitational waves (GWs) has opened a new window into astronomy, giving us a unique look at merging compact objects and the behavior of extreme gravity~\cite{LIGOScientific:2016aoc,2023PhRvX..13a1048A}. Today's large laser interferometers are most sensitive to frequencies in the hertz-to-kilohertz range. However, many models of the early Universe and new physics predict gravitational waves at much higher frequencies, between $10^4$ and $10^9$ Hz. These ultrahigh-frequency (UHF) signals could be produced by a variety of sources: small primordial black holes spiraling into each other, the movement of cosmic strings, clouds of exotic particles (axions) gathering around spinning black holes, or phase transitions in the early Universe~\cite{2013PhRvL.110g1105A, 2025LRR....28...10A, Gehrman2023, Brito2020}. Detecting these high-frequency waves is a major experimental challenge. The sensitivity of standard kilometer-scale detectors drops quickly above a few kilohertz because their arms are too long to properly track such fast-changing signals, and they become limited by quantum noise in their lasers~\cite{2014PhRvD..90j2005G,PhysRevD.108.L061303}.

This gap in our observational reach has sparked strong interest in developing new quantum sensors designed for high frequencies~\cite{PhysRevLett.132.190001}. Atomic, molecular, and optical (AMO) systems are excellent candidates for laboratory-scale detectors because they maintain highly stable quantum states, can be precisely tuned to specific frequencies, and are naturally isolated from the everyday ground vibrations that disrupt larger instruments~\cite{Ludlow:2015,Bothwell:2022,Graham:2013gfa}. Recent breakthroughs in large-scale atom interferometry~\cite{Baynham2025}, alongside rapid improvements in using Rydberg atoms to measure microwaves~\cite{Sedlacek2012}, highlight just how capable these atomic systems are as sensitive, localized probes for rapidly changing fields.

A fundamental theoretical question underpinning these quantum detection efforts is how a dynamical spacetime background modifies the interaction between an excited atomic quantum system and the electromagnetic vacuum. A recent study addressed this using a scalar-field model of light~\cite{Paczos2026GWImprints}, demonstrating that a dynamical spacetime can leave potentially observable, direction-dependent signatures in the spontaneous-emission spectrum. Related analyses based on Unruh--DeWitt detectors have similarly investigated scalar-field transition rates in gravitational-wave backgrounds~\cite{PhysRevD.105.024053,Prokopec_2023}. However, scalar-field models intrinsically fail to capture the vector character of the electromagnetic field. Because gravitational waves are spin-2 tensor perturbations, their coupling to a quantum emitter is highly dependent on photon polarization and the spatial orientation of the dipole--field interaction. A full treatment based on a quantized vector electromagnetic field in a classical gravitational-wave background is therefore strictly necessary to accurately describe the resulting physical observables.

In this work, we develop a comprehensive quantum electrodynamical (QED) framework for atomic spontaneous emission in a weak, prescribed classical gravitational-wave background. For a monochromatic gravitational wave, we show that the periodic metric perturbation dynamically phase-modulates the electromagnetic vacuum modes, generating a Floquet ladder of photonic sidebands separated by the GW frequency. While the total spontaneous-emission rate integrated over all photon frequencies and emission directions remains unchanged at first order in the GW metric perturbation, the differential emission pattern exhibits a characteristic quadrupolar redistribution that directly reflects the tensor (helicity-2) symmetry of the gravitational wave. Furthermore, we formulate an open quantum system cavity-QED model to assess the experimental observability of this phenomenon. We propose a differential readout protocol based on two orthogonally oriented high-finesse microcavities, utilizing either trapped highly charged ions or Rydberg-atom ensembles, which maps the vacuum anisotropy into a measurable photon count asymmetry.

This paper is organized as follows. In Sec.~II, we quantize the electromagnetic field in a linearized plane gravitational-wave background and derive the Floquet-modulated field operators. In Sec.~III, we calculate the transient spontaneous emission amplitudes and the resulting spectral asymmetries. In Sec.~IV, we evaluate the exact angular redistribution of the emitted radiation. Section~V extends this free-space formalism into the cavity-QED regime, deriving the time-dependent Purcell decay rates. In Sec.~VI, we outline the proposed cross-cavity experimental architecture and evaluate its fundamental strain sensitivity limits. Concluding remarks are provided in Sec.~VII.

\section{Quantization of the Electromagnetic Field in a GW Background}

In this section, we construct the quantized electromagnetic field in a dynamic spacetime. Specifically, we want to understand how electromagnetic vacuum modes, characterized by an angular frequency $\omega_k = c|\mathbf{k}|$, interact with a passing, monochromatic gravitational wave of angular frequency $\Omega$. 

We begin by considering a weak plane gravitational wave propagating along the positive $z$-direction on a flat Minkowski background. The spacetime metric is written as
\begin{equation}
g_{\mu\nu}=\eta_{\mu\nu}+h_{\mu\nu},
\qquad
|h_{\mu\nu}|\ll 1,
\end{equation}
where $x^\mu=(ct,\mathbf{x})$ are the spacetime coordinates and $\eta_{\mu\nu}=\mathrm{diag}(-1,1,1,1)$ is the unperturbed Minkowski metric. It is most convenient to work in the transverse-traceless (TT) gauge, where the metric perturbation $h_{\mu\nu}$ satisfies
\begin{equation}
h_{0\mu}=0,
\qquad
h^i_{\ i}=0,
\qquad
\partial_j h^{ij}=0.
\end{equation}
In this gauge, the spatial components of the metric perturbation can be decomposed into two independent polarization states, typically called ``plus'' ($+$) and ``cross'' ($\times$):
\begin{equation}
h_{ij}(u)
=
h_+(u)e^+_{ij}
+
h_\times(u)e^\times_{ij},
\end{equation}
where $u=t-z/c$ is the retarded time. The polarization tensors are defined using the Cartesian unit vectors $\hat{x}_i$ and $\hat{y}_i$ as
\begin{equation}
e^+_{ij}
=
\hat{x}_i\hat{x}_j-\hat{y}_i\hat{y}_j,
\qquad
e^\times_{ij}
=
\hat{x}_i\hat{y}_j+\hat{y}_i\hat{x}_j.
\end{equation}

To see how this gravitational background affects the electromagnetic field, we start with Maxwell's equations in curved spacetime. In the absence of free charges and currents, these take the covariant form~\cite{Birrell1982,Wald:1995yp}
\begin{equation}
\nabla_\mu F^{\mu\nu}
=
\frac{1}{\sqrt{-g}}
\partial_\mu
\left(
\sqrt{-g}\,
g^{\mu\alpha}g^{\nu\beta}F_{\alpha\beta}
\right)
=0,
\label{eq:Maxwell_Curved}
\end{equation}
where $F_{\mu\nu} = \partial_\mu A_\nu-\partial_\nu A_\mu$ is the electromagnetic field tensor constructed from the vector potential $A_\mu$, and $g\equiv\det(g_{\mu\nu}) = -1+\mathcal{O}(h^2)$ is the determinant of the metric. 

By adopting the generalized Lorenz gauge condition, $\nabla_\mu A^\mu=0$, the wave equation for the vector potential simplifies to
\begin{equation}
\nabla_\alpha\nabla^\alpha A^\nu
-
R^\nu_{\ \mu}A^\mu
=0,
\label{eq:vector_wave_curved}
\end{equation}
where $R_{\mu\nu}$ is the Ricci curvature tensor. Because our gravitational-wave background is a vacuum solution to Einstein's equations, it is Ricci-flat to first order in the strain ($R_{\mu\nu}=0$). Therefore, the curvature coupling term in Eq.~\eqref{eq:vector_wave_curved} vanishes entirely.

To solve this, we focus on the leading eikonal approximation. In our regime of interest, the electromagnetic frequency is vastly higher than the gravitational wave frequency ($\omega_k\gg\Omega$). Because the light oscillates so rapidly compared to the slow oscillation of spacetime, we can safely neglect small corrections to the wave's amplitude and polarization. Under these conditions, the spatial components of the vector potential satisfy a simplified wave equation:
\begin{equation}
\left(
\frac{\partial^2}{\partial t^2}
-c^2\nabla^2
\right)
A_i(t,\mathbf{x})
\simeq
-c^2h^{jk}(t,z)
\partial_j\partial_k A_i(t,\mathbf{x}).
\label{eq:wave_perturbed}
\end{equation}

From the perspective of a single atom, the gravitational wave has an enormous wavelength. We can therefore treat the gravitational field as essentially uniform across the entire atomic wavefunction, allowing us to approximate the local metric perturbation near the atom as $h_{ij}(t,z)\simeq h_{ij}(t,0)$.

For an electromagnetic mode with wave vector $\mathbf{k}$, we adopt an eikonal ansatz where the wave's phase accumulates over time:
\begin{equation}
\label{A}
A_i(t,\mathbf{x}) = \mathcal{A}_i \exp\left[ i\mathbf{k}\cdot\mathbf{x} -iS_{\mathbf{k}}(t) \right].
\end{equation}
Here, we treat the amplitude $\mathcal{A}_i$ as constant, while all the dynamical effects of the curved spacetime are absorbed into the phase $S_{\mathbf{k}}(t)$. Substituting this ansatz into Eq.~\eqref{eq:wave_perturbed} and keeping only the dominant terms ($\dot{S}_{\mathbf{k}}^2 \gg \ddot{S}_{\mathbf{k}}$) yields an instantaneous angular frequency:
\begin{align}
\dot{S}_{\mathbf{k}}(t)
&=
\omega_k+\delta\omega_{\mathbf{k}}(t)
\nonumber\\
&=
ck
-\frac{c}{2k}
h^{ij}(t,0)k_i k_j
\nonumber\\
&=
\omega_k
\left[
1-\frac{1}{2}
h^{ij}(t,0)\hat{k}_i\hat{k}_j
\right],
\label{eq:instantaneous_frequency}
\end{align}
where $\omega_k=ck$ and $\hat{\mathbf{k}}=\mathbf{k}/k$. This result makes intuitive sense: as the gravitational wave stretches and squeezes the local geometry, it continuously shifts the effective frequency of the passing light.

To see how this phase accumulates, let us consider a simple monochromatic, linearly polarized gravitational wave. Assuming both polarization components share a common phase, we write:
\begin{equation}
h_{ij}(t,0)
=
A_{ij}
\cos\left(\Omega t+\varphi_{\mathrm{GW}}\right),
\qquad
A_{ij}
=
A_+e^+_{ij}+A_\times e^\times_{ij}.
\end{equation}
Integrating the instantaneous frequency from Eq.~\eqref{eq:instantaneous_frequency} gives the total accumulated phase:
\begin{equation}
S_{\mathbf{k}}(t)
=
\omega_k t
-
C_{\mathbf{k}}
\sin\left(\Omega t+\varphi_{\mathrm{GW}}\right),
\label{eq:phase_modulated}
\end{equation}
where the dimensionless modulation index $C_{\mathbf{k}}$ captures the strength of the interaction between the light and the gravitational wave:
\begin{equation}
C_{\mathbf{k}}
=
\frac{\omega_k}{2\Omega}
\left[
A_+\left(\hat{k}_x^2-\hat{k}_y^2\right)
+
2A_\times\hat{k}_x\hat{k}_y
\right].
\label{eq:ModulationIndex}
\end{equation}
Notice that $|C_{\mathbf{k}}|\simeq \frac{h_0}{2}\frac{\omega_k}{\Omega}$. To get a sense of scale, let us plug in some typical 
numbers for a range of gravitational wave interacting with an electromagnetic wave in the optical wavelength:
\begin{equation}
\label{estimation}
\frac{\omega_k}{2\pi}\sim10^{15}\,\mathrm{Hz},
\qquad
\frac{\Omega}{2\pi}\sim10^4\text{--}10^9\,\mathrm{Hz},
\qquad
h_0\sim10^{-20}.
\end{equation}
This gives a modulation index of approximately
\begin{equation}
|C_{\mathbf{k}}|
\simeq
5\times10^{-10}~\text{--}~5\times10^{-15}.
\end{equation}
While tiny, this represents a cumulative, physical phase shift imprinted on the electromagnetic wave by the dynamic spacetime. 

Having established how a classical electromagnetic wave behaves in this background, we must now quantize the electromagnetic field to study its interaction with quantum emitters. We treat gravity semiclassically: the gravitational wave remains a prescribed classical background metric, while the electromagnetic field is promoted to a quantum operator. We achieve this by taking our classical eikonal solutions and promoting the constant amplitudes $\mathcal{A}_i$ to the standard quantum annihilation and creation operators ($\hat{a}_{\mathbf{k}\lambda}$ and $\hat{a}^{\dagger}_{\mathbf{k}\lambda}$), which obey the canonical commutation relations:
\begin{equation}
\left[
\hat{a}_{\mathbf{k}\lambda},
\hat{a}^{\dagger}_{\mathbf{k}'\lambda'}
\right]
=
\delta_{\mathbf{k}\mathbf{k}'}
\delta_{\lambda\lambda'}.
\end{equation}
The classical phase modulation $S_{\mathbf{k}}(t)$ is retained in the mode functions. To express this oscillating phase in a form that is easier to evaluate, we apply the Jacobi--Anger expansion:
\begin{equation}
e^{iC_{\mathbf{k}}\sin\theta}
=
\sum_{n=-\infty}^{\infty}
J_n(C_{\mathbf{k}})e^{in\theta},
\end{equation}
where $J_n$ denotes the Bessel function of the first kind. By combining the canonical operators with this expanded phase, the fully quantized vector potential operator becomes:
\begin{align}
\hat{\mathbf{A}}(t,\mathbf{x})
={}&
\sum_{\mathbf{k},\lambda}
\sqrt{
\frac{\hbar}
{2\epsilon_0\omega_k V}
}
\,
\boldsymbol{\epsilon}_{\mathbf{k}\lambda}
\hat{a}_{\mathbf{k}\lambda}
e^{i\mathbf{k}\cdot\mathbf{x}}
\nonumber\\
&\times
\sum_{n=-\infty}^{\infty}
J_n(C_{\mathbf{k}})
e^{-i(\omega_k-n\Omega)t
+in\varphi_{\mathrm{GW}}}
+\mathrm{H.c.}
\label{eq:Quantized_Field_Floquet}
\end{align}
Here, we use hats to denote operators. The vector $\boldsymbol{\epsilon}_{\mathbf{k}\lambda}$ is the transverse polarization vector (satisfying $\mathbf{k}\cdot\boldsymbol{\epsilon}_{\mathbf{k}\lambda}=0$) and $V$ is the local quantization volume. 

What Equation~\eqref{eq:Quantized_Field_Floquet} tells us is physically quite intuitive. Instead of simply nudging the electromagnetic vacuum modes to a new frequency, the rippling spacetime fundamentally reshapes---or ``dresses''---them. It splits each original mode into a ladder of new frequencies, with every rung spaced apart by exactly the gravitational wave frequency $\Omega$. The most important detail here is that these rungs are not separate, independent photons. They are all tied together as a single package, linked to the same underlying quantum state (the annihilation operator $\hat{a}_{\mathbf{k}\lambda}$). In the next section, we will see what happens when an atom spontaneously emits a photon into this intricately patterned vacuum.

\section{Spontaneous Emission}

To see how this dressed vacuum affects light-matter interactions, let's model our emitting atom as a simple two-level system. It has an excited state $|e\rangle$, a ground state $|g\rangle$, a natural transition frequency $\omega_0$, and an electric dipole matrix element $\mathbf{d} = \langle g | \hat{\mathbf{d}} | e \rangle$. Using the standard electric dipole approximation, the interaction between the atom and the vacuum field is governed by the Hamiltonian
\begin{equation}
\label{hamiltonian}
\hat{H}_I(t) = -\hat{\mathbf{d}} \cdot \hat{\mathbf{E}}(t, \mathbf{0}),
\end{equation}
where the electric field at the atom's position is simply $\hat{\mathbf{E}}(t, \mathbf{0}) = -\partial \hat{\mathbf{A}}/\partial t$.

Suppose the atom starts in its excited state $|e, 0\rangle$ at some initial time $\tau_i$. We want to find the probability amplitude of it emitting a photon into a specific mode $(\mathbf{k}, \lambda)$ after interacting with the field for a duration $T = \tau_f - \tau_i$. Using standard first-order perturbation theory, we plug our dressed field operator into the interaction Hamiltonian and integrate over time:
\begin{align}
\beta_{\mathbf{k}\lambda}(T) &= -\frac{i}{\hbar} \int_{\tau_i}^{\tau_f} d\tau \langle g, 1_{\mathbf{k}\lambda} | \hat{H}_I(\tau) | e, 0 \rangle \nonumber \\
&= \mathcal{M}_{\mathbf{k}\lambda} \sum_{n=-\infty}^{\infty} J_n(C_\mathbf{k}) e^{-i \Delta_n \tau_c} T\, \text{sinc}\left( \frac{\Delta_n T}{2} \right),
\label{eq:Amplitude_Sinc}
\end{align}
where $\mathcal{M}_{\mathbf{k}\lambda} = \sqrt{\frac{\omega_k}{2 \epsilon_0 \hbar V}} (\mathbf{d} \cdot \mathbf{e}_{\mathbf{k}\lambda}^*)$ is the transition matrix element, $\tau_c = (\tau_i + \tau_f)/2$ is the central time of the interaction, and our effective detuning is 
\begin{equation}
\Delta_n \equiv \omega_0 - \omega_k + n \Omega.
\label{eq:Detuning_Corrected}
\end{equation}
To understand exactly where this detuning comes from, we just need to track the internal clocks of the atom and the field. The atomic dipole operator naturally ticks at its transition frequency, evolving as $e^{-i\omega_0 t}$. Meanwhile, the quantized field from Eq.~(\ref{eq:Quantized_Field_Floquet}) carries a time dependence of $e^{+i(\omega_k - n\Omega)t}$. When they interact, their combined phase evolves as $e^{-i(\omega_0 - \omega_k + n\Omega)\tau}$. Integrating this oscillating phase over the interaction window gives us the familiar sinc function, centered around the modified detuning $\Delta_n$.

Because the gravitational wave strain is incredibly weak ($|C_\mathbf{k}| \ll 1$), we can simplify our picture by keeping only the dominant terms in the Bessel expansion: $J_0(C_\mathbf{k}) \approx 1$ and $J_{\pm 1}(C_\mathbf{k}) \approx \pm C_\mathbf{k} / 2$. With this truncation, the spectral photon density $\langle n_{\mathbf{k}\lambda} \rangle = |\beta_{\mathbf{k}\lambda}|^2$ naturally splits into three distinct pieces:
\begin{equation}
\langle n_{\mathbf{k}\lambda} \rangle = |\mathcal{M}_{\mathbf{k}\lambda}|^2 T^2 \left[ \mathcal{S}_0(\delta_k) + C_\mathbf{k} \cos(\psi) \mathcal{D}_{\text{int}}(\delta_k; \Omega) + \mathcal{O}(C_\mathbf{k}^2) \right].
\label{n}
\end{equation}
Here, $\delta_k = \omega_0 - \omega_k$ is the standard free-space detuning, $\psi = \Omega \tau_c + \varphi_{\text{GW}}$ is the local phase of the gravitational wave, and $\mathcal{S}_0(\delta_k) = \text{sinc}^2(\delta_k T / 2)$ is the usual unperturbed emission profile. The exciting new physics is captured by the interference profile:
\begin{equation}
\mathcal{D}_{\text{int}} = \text{sinc}\left(\frac{\delta_k T}{2}\right) \left[ \text{sinc}\left( \frac{(\delta_k + \Omega)T}{2} \right) - \text{sinc}\left( \frac{(\delta_k - \Omega)T}{2} \right) \right].
\label{eq:D_interference_Corrected}
\end{equation}

\begin{figure}[t]
    \centering
    \includegraphics[width=0.98\linewidth]{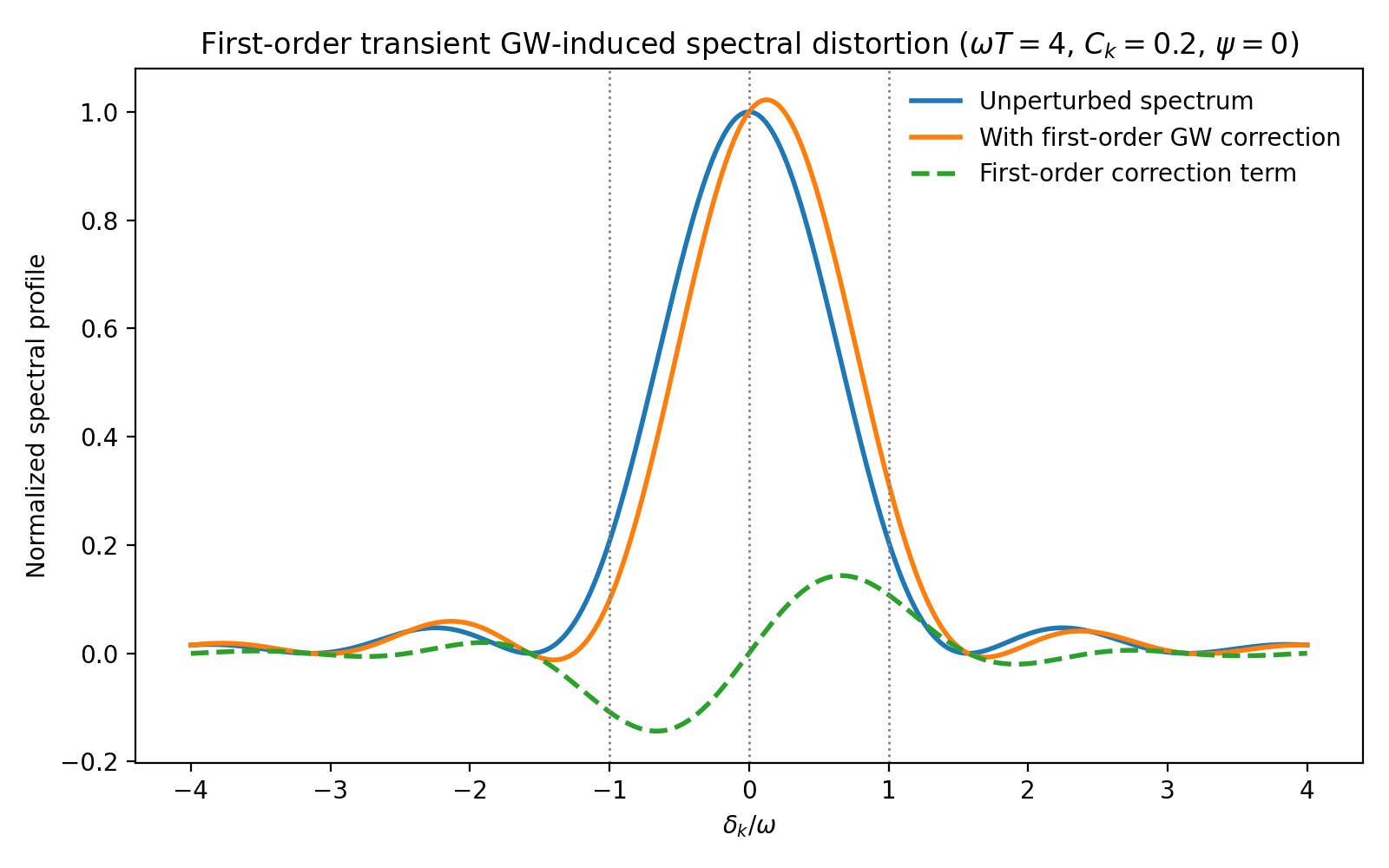}
    \caption{Transient spontaneous emission spectrum $\langle n_{\mathbf{k}\lambda} \rangle$ at a finite interaction time ($T = 4/\Omega$). The quantum interference profile $\mathcal{D}_{\text{int}}$ induces a phase-dependent asymmetry across the carrier line $\delta_k = 0$. (Note: The GW modulation index is exaggerated to $C_{\mathbf{k}} = 0.2$ here for visual clarity).}
    \label{fig:transient_spectrum}
\end{figure}

Figure~\ref{fig:transient_spectrum} shows what this transient emission spectrum looks like when the interaction time is comparable to the gravitational wave period ($T \sim \Omega^{-1}$). At this timescale, the Floquet sidebands are broad enough to heavily overlap with the central carrier frequency. Notice how the interference term $\mathcal{D}_{\text{int}}$ is an odd function (flipping sign if $\delta_k \to -\delta_k$). This creates a distinct "red-blue" spectral asymmetry—a slight tilting of the emission line that oscillates in step with the gravitational wave phase $C_\mathbf{k} \cos\psi$.

\section{Gravitational-Wave-Induced Quadrupolar Angular Redistribution}

In the previous section, we looked at the transient spectral effects of the gravitational wave. Now, let us examine the spatial pattern of the emitted radiation in the long-time limit ($T \gg \Gamma_0^{-1}$), where the emission resolves into a steady angular distribution. 

When we sum over the possible photon polarizations $\mathbf{e}_{\mathbf{k}\lambda}$ in our emission amplitude, we isolate the component of the atomic dipole that is transverse to the photon's emission direction:
\begin{equation}
\sum_{\lambda=1}^2 |\mathbf{d} \cdot \mathbf{e}_{\mathbf{k}\lambda}^*|^2 =  |\mathbf{d}_\perp|^2.
\end{equation}
Because the gravitational wave possesses a characteristic spin-2 quadrupolar geometry, it is natural to suspect that the spontaneously emitted photons will be steered into a similar quadrupolar pattern. To test this intuition, we calculate the differential decay rate per unit solid angle:
\begin{equation}
\frac{d\Gamma}{d\Omega_k} =  \int_0^\infty \frac{k^2\,dk}{(2\pi)^3}\, \frac{1}{T}\, \langle n_{\mathbf{k}\lambda}\rangle,
\label{eq:dGamma_dOmega_def}
\end{equation}
where $d\Omega_k=\sin\theta\,d\theta\,d\phi$.

We can split this rate into the standard unperturbed atomic emission plus a small correction induced by the gravitational wave:
\begin{equation}
\frac{d\Gamma}{d\Omega_k} = \left(\frac{d\Gamma}{d\Omega_k}\right)_0 + \delta\left(\frac{d\Gamma}{d\Omega_k}\right).
\end{equation}
By substituting the unperturbed part of our photon density from Eq.~\eqref{n} into Eq.~\eqref{eq:dGamma_dOmega_def}, we perfectly recover the familiar dipole emission pattern of an atom in free space:
\begin{equation}
\label{rate}
\left(\frac{d\Gamma}{d\Omega_k}\right)_0 = \frac{\omega_0^3 |\mathbf{d}_\perp|^2}{8 \pi^2 \epsilon_0 \hbar c^3}.
\end{equation}

To find the GW-induced correction, we need to evaluate the modulation index $C_\mathbf{k}$ on resonance ($\omega_k = \omega_0$). The angles seamlessly emerge when we express the photon's emission direction $\hat{\mathbf{k}}$ in spherical coordinates relative to the GW propagation axis ($z$). Using $\hat{k}_x = \sin\theta \cos\phi$ and $\hat{k}_y = \sin\theta \sin\phi$, the geometric terms in Eq.~\eqref{eq:ModulationIndex} transform via double-angle identities into $\hat{k}_x^2 - \hat{k}_y^2 = \sin^2\theta \cos(2\phi)$ and $2\hat{k}_x \hat{k}_y = \sin^2\theta \sin(2\phi)$. 

If we define $\Delta(\theta,\phi) \equiv \delta(d\Gamma/d\Omega_k) / (d\Gamma/d\Omega_k)_0$ as the fractional contrast of the anisotropic emission, these spherical coordinates give us a beautifully explicit formula for the first-order perturbation:
\begin{equation}
\Delta(\theta,\phi)=  \frac{\omega_0}{2 \Omega} \sin^2\theta 
 \left[ A_+ \cos(2\phi) + A_\times \sin(2\phi) \right] \cos\psi.
\label{eq:delta_rate_quadrupolar}
\end{equation}
Based on our earlier estimates, the amplitude of this contrast is roughly $\frac{h_0}{2}\frac{\omega_0}{\Omega}$, which corresponds to the numerical scale established in Eq.~\eqref{estimation}. 

A striking physical consequence reveals itself when we integrate this contrast over the entire solid angle:
\begin{equation}
 \int_{0}^{2\pi} d\phi \int_{0}^{\pi} d\theta \, \sin\theta \,
\Delta(\theta,\phi) = 0.
\end{equation}
Because the integrals of $\cos(2\phi)$ and $\sin(2\phi)$ over a full circle are identically zero, the total spontaneous decay rate remains completely unchanged to first order in the GW strain. The gravitational wave does not alter the overall probability of emission or the atom's lifetime; instead, it redistributes the emitted photons according to the spatial "stretching and squeezing" of the metric. 

This anisotropic redistribution---proportional to $\sin^{2}\theta\cos(2\phi)$---is the principal observable signature of gravitational-wave-modified spontaneous emission. In the following section, we will explore how a cavity-QED architecture can be designed to exploit this precise angular pattern for GW detection.

\section{Cavity QED Dynamics in a GW Field}

A cavity quantum electrodynamics (cavity-QED) setup provides the perfect environment to catch this subtle effect. It allows us to convert the weak, highly directional modifications of the electromagnetic vacuum into a measurable optical or microwave signal, supercharging the emission into specific directions via the Purcell effect~\cite{Purcell1946,Haroche2006,ScullyZubairy1997}.

In free space, as we just saw, a passing gravitational wave leaves the total spontaneous decay rate completely unchanged; its only footprint is the quadrupolar shuffling of emitted photons. If we tried to measure this in open space, the signal would be impossibly diluted over $4\pi$ steradians. A high-finesse cavity solves this limitation beautifully in three ways. First, the Purcell effect acts like a funnel, concentrating a large fraction of the atomic decay into a single resonant mode and massively amplifying the absolute photon flux carrying the GW signature. Second, by carefully aiming the cavity axis (which we will call $\hat{\mathbf{k}}_c$), we can deliberately sample the emission at the exact angle where the quadrupolar contrast is strongest ($\theta_c=\pi/2$), avoiding the GW propagation axis ($\theta_c=0$) where the modulation vanishes. Third, using a pair of orthogonally oriented cavities allows for a differential readout. For a $+$-polarized GW, a cavity aligned along the $x$-axis ($\phi_c=0$) will see an enhanced emission rate, while a cavity along the $y$-axis ($\phi_c=\pi/2$) sees a suppressed rate. Subtracting these two outputs isolates the pure tensor GW response while canceling out everyday common-mode technical noise.

To make this quantitative, imagine enclosing our atomic emitter inside a rigid, high-finesse cavity tuned to the atom's transition frequency ($\omega_c \approx \omega_0$). The cavity selects a single spatial mode along $\hat{\mathbf{k}}_c$. How strongly the atom "talks" to this mode is governed by a coupling parameter known as the vacuum Rabi frequency, $g(t)$.  To be precise, the vacuum Rabi frequency has the dimension of angular frequency and physically, it represents the rate at which the atom and the cavity vacuum exchange a single quantum of energy. In a quiet, unperturbed vacuum, this coupling is a constant $g_0 = (\mathbf{d} \cdot \mathbf{e}_\lambda) \sqrt{\omega_c / (2 \epsilon_0 \hbar V_c)}$, which depends entirely on the strength of the atomic dipole $\mathbf{d}$ and how tightly the cavity pinches the electric field (the mode volume $V_c$). 

Because our physical cavity is incredibly rigid and microscopic compared to the kilometer-scale wavelength of the gravitational wave, its geometry and basic resonance $\omega_c$ are unaffected by the passing GW. However, the GW  dress the local electric field. If we evaluate the dressed electric field operator $\hat{\mathbf{E}}_{\text{cav}}(t,\mathbf{0})=-\partial_t\hat{\mathbf{A}}_{\text{cav}}(t,\mathbf{0})$ at the atom's position and plug it into our interaction Hamiltonian, we find that the effective atom-cavity coupling acquires a tiny time-dependent ripple:
\begin{equation}
g(t) = g_0 \left[ 1 - \frac{C_{\mathbf{k}_c}\,\Omega}{\omega_c} \cos(\Omega t + \varphi_{\text{GW}}) \right].
\label{eq:TimeDependentRabi}
\end{equation}
It is important to notice the scale of this direct coupling modification. Because our modulation index is $C_{\mathbf{k}_c} \simeq \frac{h_0}{2}\frac{\omega_c}{\Omega}$, the relative change in the Rabi frequency is roughly $\delta g/g_0 \sim h_0$. This is vanishingly small. It is completely overshadowed by the angular redistribution contrast $\Delta \sim C_{\mathbf{k}_c}$ we derived earlier. 

So, how does this translate into the actual rate of photons spilling into the cavity? We calculate this using Fermi's Golden Rule~\cite{Sakurai2020}. For an excited two-level quantum system, first-order time-dependent perturbation theory yields the transition rate from an initial state $|i\rangle$ to a continuum of final states $|f\rangle$:
\begin{equation}
\Gamma = \frac{2\pi}{\hbar}\,|\mathcal{M}_{fi}|^2\,\rho(\omega_0),
\label{eq:FermiGoldenRule}
\end{equation}
where $\mathcal{M}_{fi}=\langle f|\hat{H}_I|i\rangle$ is the interaction matrix element between the states, and $\rho(\omega_0)$ is the density of final states per unit energy, evaluated right at the transition frequency. In free space, the continuum of electromagnetic modes provides the standard photon density of states $\rho_{\rm free}(\omega)=\omega^2 V/(\pi^2\hbar c^3)$~\cite{ScullyZubairy1997,Milonni1994}, giving us the familiar baseline spontaneous emission rate $\Gamma_0=\omega_0^3|\mathbf{d}|^2/(3\pi\epsilon_0\hbar c^3)$. Inside a cavity, however, the rigid mirror boundaries slice up the spectrum, squeezing the density of states into a sharp, narrow resonance. In the following subsection, we will step through exactly how this altered density of states transforms the overall emission rate.

\subsection{Rate of spontaneous emission in cavity}
For the single cavity mode of interest, the interaction Hamiltonian in the electric-dipole approximation is $\hat{H}_I = -\hat{\mathbf{d}}\cdot\hat{\mathbf{E}}_{\rm cav}$. The cavity electric field at the atomic position ($\mathbf{x}=\mathbf{0}$) is
\begin{equation}
\hat{\mathbf{E}}_{\rm cav}(\mathbf{0}) = i\,\mathbf{e}_\lambda\,
\sqrt{\frac{\hbar\omega_c}{2\epsilon_0 V_c}}\,
(\hat{a}-\hat{a}^\dagger),
\end{equation}
where $\mathbf{e}_\lambda$ is the polarization vector, $V_c$ is the cavity mode volume, and $\hat{a}$ is the annihilation operator for the resonant mode. The atomic dipole operator between the excited and ground states is $\hat{\mathbf{d}} = \mathbf{d}\,\hat{\sigma}_+ + \mathbf{d}^*\hat{\sigma}_-$. Taking the initial state $|i\rangle=|e,0_c\rangle$ (atom excited, cavity empty) and the final state $|f\rangle=|g,1_c\rangle$ (atom ground, one photon in the cavity), the matrix element is
\begin{equation}
\mathcal{M}_{fi} = \langle g,1_c|\hat{H}_I|e,0_c\rangle
= \hbar g_0,
\qquad
g_0 = (\mathbf{d}\cdot\mathbf{e}_\lambda)\,
\sqrt{\frac{\omega_c}{2\epsilon_0\hbar V_c}}.
\end{equation}
Hence $|\mathcal{M}_{fi}|^2 = \hbar^2 g_0^2$, where $g_0$ is the vacuum Rabi frequency introduced in Eq.~\eqref{eq:TimeDependentRabi}.

A realistic cavity is not perfectly closed; photons leak out through the partially transmitting mirrors at a rate $\kappa$. This loss broadens the discrete cavity resonance into a Lorentzian lineshape. To see this, consider the classical amplitude $a(t)$ of the cavity field. In the absence of driving, it obeys $\dot{a} = -(\kappa/2)a$, so the energy $a$ decays as $e^{-\kappa/2 t}$. The Fourier transform of standing wave and the exponential decay (i.e. $e^{-\kappa/2 t} e^{-i\omega_c t}$) gives a spectral intensity proportional to $[(\omega-\omega_c)^2+(\kappa/2)^2]^{-1}$. Normalizing such that $\int\rho_{\rm cav}(E)\,dE = 1$ (one mode), the density of states per unit energy is
\begin{equation}
\rho_{\rm cav}(E) = \frac{1}{2\pi}\,
\frac{\hbar\kappa}{(E-\hbar\omega_c)^2+(\hbar\kappa/2)^2}.
\end{equation}
Using $E=\hbar\omega$ and $\rho(\omega)=\hbar\rho(E)$, the density of states per unit angular frequency is
\begin{equation}
\rho_{\rm cav}(\omega) = \frac{1}{2\pi}\,
\frac{\kappa}{(\omega-\omega_c)^2+(\kappa/2)^2},
\qquad
\kappa \equiv \frac{\omega_c}{Q},
\label{eq:CavityDensityOfStates}
\end{equation}
where $Q$ is the quality factor of the cavity. On exact resonance ($\omega_0=\omega_c$), this evaluates to
\begin{equation}
\rho_{\rm cav}(\omega_0) = \frac{2}{\pi\kappa}.
\label{eq:ResonantDensity}
\end{equation}
The narrower the linewidth (larger $Q$), the higher the peak density of states, and the stronger the Purcell enhancement.

Now we substitute $|\mathcal{M}_{fi}|^2=\hbar^2 g_0^2$ and Eq.~\eqref{eq:ResonantDensity} into Fermi's Golden Rule \eqref{eq:FermiGoldenRule} yields the spontaneous emission rate into the cavity mode:
\begin{equation}
\Gamma_{\rm cav,0} = \frac{2\pi}{\hbar^2}\,(\hbar^2 g_0^2)\,
\left(\frac{2}{\pi\kappa}\right)
= \frac{4g_0^2}{\kappa}.
\label{eq:PurcellRateBare}
\end{equation}
This is the standard Purcell result in the weak-coupling regime. It can be rewritten as
\begin{equation}
\Gamma_{\rm cav,0} = F_P\,\Gamma_0,
\qquad
F_P = \frac{4g_0^2}{\kappa\Gamma_0},
\label{eq:PurcellFactor}
\end{equation}
where $\Gamma_0$ is the free-space spontaneous emission rate and $F_P$ is the Purcell enhancement factor. For a high-finesse microcavity, $F_P$ can range from order unity to $10^4$ or more, effectively concentrating a large fraction of the atomic decay into the resonant mode.

Finally we come to the question that how the GW modifies Eq.~\eqref{eq:PurcellRateBare}. As argued above, the cavity mirrors are rigid and the cavity size is much smaller than the GW wavelength. Consequently,  the cavity proper length, and the bare vacuum Rabi frequency $g_0$ are all unperturbed to leading order. The cavity linewidth $\kappa$ is likewise unaffected because it is determined by the mirror reflectivity and the fixed cavity geometry.

The only quantity that the GW modifies is the \emph{angular distribution} of the electromagnetic vacuum. In free space, the differential spontaneous emission rate into direction $\hat{\mathbf{k}}$ carries the quadrupolar contrast $\Delta(\theta,\phi)$ derived in Eq.~\eqref{eq:delta_rate_quadrupolar}. Because a single-mode cavity samples emission into a single direction $\hat{\mathbf{k}}_c$, the effective transition rate into that mode inherits the same angular contrast according to Eq. (\ref{eq:delta_rate_quadrupolar}) is:
\begin{equation}
\Gamma_{\rm cav}(t) = \Gamma_{\rm cav,0}\,
\bigl[1+\Delta(\theta_c,\phi_c)\bigr].
\label{eq:RateWithContrast}
\end{equation}
This is the crucial step: the cavity converts the free-space angular anisotropy into a time-dependent decay rate. Using Eq.~\eqref{eq:delta_rate_quadrupolar} with $\Delta(\theta_c,\phi_c)=C_{\mathbf{k}_c}\cos(\Omega t+\varphi_{\rm GW})$ and substituting Eqs.~\eqref{eq:PurcellRateBare}--\eqref{eq:PurcellFactor} into Eq.~\eqref{eq:RateWithContrast}, we obtain
\begin{align}
\Gamma_{\rm cav}(t) &= \frac{4g_0^2}{\kappa}
\left[1 + C_{\mathbf{k}_c}\cos(\Omega t+\varphi_{\rm GW})\right]
\nonumber\\[4pt]
&= F_P\,\Gamma_0
\left[1 + C_{\mathbf{k}_c}\cos(\Omega t+\varphi_{\rm GW})\right].
\label{eq:PurcellRateModulated}
\end{align}

Equation~\eqref{eq:PurcellRateModulated} is the central result of this section. It shows that the cavity converts the angular anisotropy of the GW-modified vacuum into a time-dependent atomic decay rate. The modulation amplitude is set by the directional index $C_{\mathbf{k}_c}$, which vanishes when the cavity axis is aligned with the GW propagation direction ($\theta_c=0$) and is maximized in the transverse plane ($\theta_c=\pi/2$). The Purcell factor $F_P$ amplifies the absolute photon flux, while the quadrupolar angular dependence $\Delta(\theta_c,\phi_c)$ encodes the spin-2 tensor character of the gravitational wave. In the next section we examine how a differential cross-cavity measurement can isolate this signal from technical noise.

\section{Experimental Architecture and Sensitivity Limits}
\label{sec:experiment}

\begin{figure}[htbp]
    \centering
    \begin{tikzpicture}[>=Stealth, scale=1.15]
        \draw[->, dashed, gray!70] (-3.2,0) -- (3.2,0) node[right] {$x$};
        \draw[->, dashed, gray!70] (0,-2.5) -- (0,3.2) node[above] {$y$};
        \draw[->, dashed, gray!70] (0,0) -- (-2.2,-2.2) node[below left] {$z$};

        \draw[thick, gray!50, rounded corners=4pt] (-1.8,-1.8) rectangle (1.8,1.8);
        \node[gray!70, font=\tiny] at (1.1, -1.6) {Rigid Substrate};

        \shade[ball color=blue!40] (0,0) circle (0.35cm);
        \node[below right, text width=2cm, align=left] at (0.2, -0.2) {\small Emitters \\[-0.5ex] \footnotesize (Rydberg / HCI)};

        \draw[ultra thick, black] (1.5, -0.6) arc (-30:30:1.2cm);
        \draw[ultra thick, black] (-1.5, 0.6) arc (150:210:1.2cm);

        \draw[ultra thick, black] (-0.6, 1.5) arc (120:60:1.2cm);
        \draw[ultra thick, black] (0.6, -1.5) arc (-60:-120:1.2cm);

        \node[draw, thick, fill=green!10, minimum width=0.8cm, minimum height=0.7cm, rounded corners=1pt] (Dx) at (2.8, 0) {$D_x$};
        \node[draw, thick, fill=green!10, minimum width=0.7cm, minimum height=0.8cm, rounded corners=1pt] (Dy) at (0, 2.8) {$D_y$};

        \draw[decorate, decoration={snake, segment length=2.5mm, amplitude=0.6mm}, ->, red, thick] (1.65, 0) -- (Dx.west) node[midway, above=1mm] {$\Gamma_{\text{cav}}^{(x)}$};
        \draw[decorate, decoration={snake, segment length=2.5mm, amplitude=0.6mm}, ->, red, thick] (0, 1.65) -- (Dy.south) node[midway, right=1mm] {$\Gamma_{\text{cav}}^{(y)}$};

        \draw[decorate, decoration={snake, segment length=3.5mm, amplitude=1mm}, ->, purple, ultra thick] (-2.9, -2.9) -- (-0.8, -0.8);
        \node[purple, font=\bfseries, align=center] at (-3.4, -3.2) {GW ($h_+$) \\[-0.5ex] $\mathbf{k}_{\text{GW}} \parallel \hat{\mathbf{z}}$};

        \draw[->, very thick, blue!80!black] (0.4, 0.15) -- (1.1, 0.15);
        \draw[->, very thick, blue!80!black] (-0.4, 0.15) -- (-1.1, 0.15);
        \node[blue!80!black, font=\small] at (0.8, 0.35) {$\delta \Gamma_{\text{cav}}^{(x)} > 0$};

        \draw[->, very thick, red!80!black] (0.15, 0.4) -- (0.15, 1.1);
        \draw[->, very thick, red!80!black] (0.15, -0.4) -- (0.15, -1.1);
        \node[red!80!black, font=\small] at (-0.6, 0.8) {$\delta \Gamma_{\text{cav}}^{(y)} < 0$};
    \end{tikzpicture}
    \caption{Schematic of the differential cross-cavity QED architecture. An atomic ensemble is trapped at the intersection of two orthogonal microcavities mounted on a rigid low-expansion spacer. A normally incident gravitational wave ($h_+$ polarization propagating along $z$) modulates the local vacuum mode density, enhancing Purcell emission into the $x$-cavity ($\delta\Gamma_{\text{cav}}^{(x)} > 0$) while suppressing it along the $y$-cavity ($\delta\Gamma_{\text{cav}}^{(y)} < 0$). Single-photon detectors $D_x$ and $D_y$ register the differential output to extract the asymmetry $\mathcal{A}_{\text{obs}}(t)$.}
    \label{fig:cavity_setup}
\end{figure}
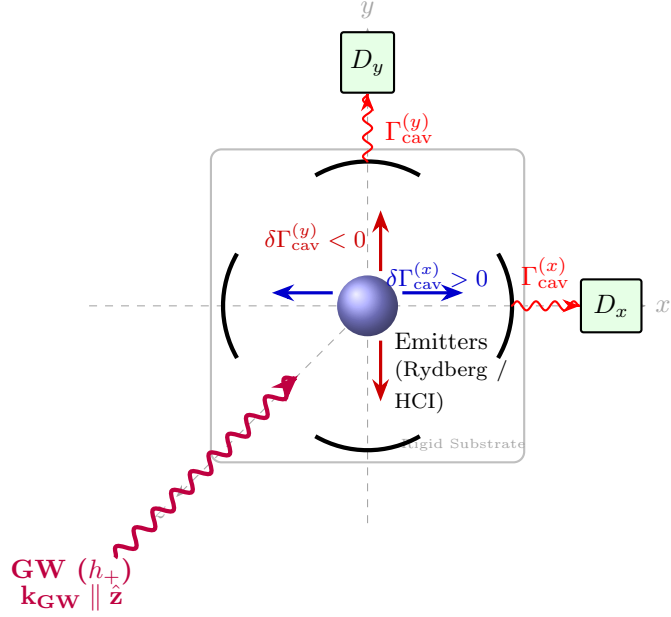

To isolate the GW-induced quadrupolar signal from isotropic background decay and thermal fluctuations, we propose the differential cross-cavity architecture illustrated in Fig.~\ref{fig:cavity_setup}. Two identical, orthogonal high-finesse microcavities are centered on a common atomic trap and aligned along the Cartesian $\hat{\mathbf{x}}$ and $\hat{\mathbf{y}}$ axes within a rigid, low-expansion substrate.

For an $h_+$ polarized GW propagating along the $\hat{\mathbf{z}}$-axis, evaluating Eq.~\eqref{eq:ModulationIndex} along the respective cavity axes yields modulation parameters of equal magnitude and opposite sign: $C_{\mathbf{k}_x} = +C_{\mathbf{k}_c}$ and $C_{\mathbf{k}_y} = -C_{\mathbf{k}_c}$. The primary experimental observable is the normalized differential photon count asymmetry,
\begin{equation}
\mathcal{A}_{\text{obs}}(t) \equiv \frac{N_x(t) - N_y(t)}{N_x(t) + N_y(t)} = C_{\mathbf{k}_c} \cos(\Omega t + \varphi_{\text{GW}}),
\label{eq:ObservableAsymmetry}
\end{equation}
where $N_x(t)$ and $N_y(t)$ are the instantaneous photon counts recorded at detectors $D_x$ and $D_y$. By constructing this balanced differential observable, common-mode cavity intensity fluctuations, atomic excitation laser noise, and unguided isotropic free-space decay channels cancel out to leading order.

The detection sensitivity depends on how the atomic transition frequencies and dipole matrix elements scale across candidate atomic systems. For hydrogenic states parameterized by nuclear charge $Z$ and principal quantum number $n$, standard atomic scaling laws give $\omega_0 \propto Z^2/n^3$, $|\mathbf{d}| \propto n^2/Z$, and unperturbed radiative decay $\Gamma_0 \propto Z^4/n^5$~\cite{Gallagher:1994}. Because the modulation index scales as $C_{\mathbf{k}_c} \propto (\omega_0/\Omega) h$, the amplitude of the modulated differential Purcell photon emission rate $\delta\Gamma \equiv F_P \Gamma_0 C_{\mathbf{k}_c}$ scales as
\begin{equation}
\delta\Gamma \propto F_P \left(\frac{\omega_0}{\Omega}h\right) \Gamma_0 \propto F_P \frac{Z^6}{n^8 \Omega} h.
\label{eq:Scaling_Z_n}
\end{equation}

\begin{table}[t]
\caption{Physical parameters and projected strain sensitivity $h_{\min}$ for Alkali Rydberg ensembles ($^{87}\text{Rb}$) and Highly Charged Ions ($^{207}\text{Pb}^{81+}$) in the cross-cavity configuration at benchmark frequencies $\Omega/2\pi = 100\text{ kHz}$ and $10\text{ MHz}$, assuming an overall detection efficiency $\eta_{\text{eff}} = 0.70$.}
\label{tab:exp_params}
\setlength{\tabcolsep}{3pt}
\begin{tabular}{lcc}
\toprule
\textbf{Parameter} & \textbf{Rydberg ($^{87}\text{Rb}$)} & \textbf{HCI($\text{Pb}^{81+})$} \\
\midrule
Principal quantum number $n$ & $45$ & $1$ \\
Nuclear charge $Z$ & $1$ & $81$ \\
Transition frequency $\omega_0 / 2\pi$ & $15.2\text{ GHz}$ & $2.6\text{ PHz}$ \\
Dipole moment $d$ ($e a_0$) & $1.2 \times 10^3$ & $1.2 \times 10^{-2}$ \\
Free-space decay $\Gamma_0$ ($\text{s}^{-1}$) & $1.4 \times 10^4$ & $3.8 \times 10^{8}$ \\
Purcell factor $F_P$ & $10^4$ & $10^4$ \\
Ensemble size $N_{\text{at}}$ & $3 \times 10^8$ & $10^4$ \\
Coherence time $T_{\text{coh}}$ & $100\text{ }\mu\text{s}$ & $10\text{ ms}$ \\
Overall efficiency $\eta_{\text{eff}}$ & $0.70$ & $0.70$ \\
Benchmark GW frequency $\Omega/2\pi$ & $100\text{ kHz}$ & $10\text{ MHz}$ \\
\midrule
\textbf{Strain sensitivity $h_{\min}$ ($1\text{ s}$ int.)} & $\mathbf{3.8 \times 10^{-14}}$ & $\mathbf{2.1 \times 10^{-17}}$ \\
\bottomrule
\end{tabular}
\end{table}

This strong $Z^6/n^8$ scaling naturally identifies two distinct operational domains. Highly Charged Ions (HCIs), such as $^{207}\text{Pb}^{81+}$ ($Z=81, n=1$), possess extreme ultraviolet and optical transitions ($\omega_0/2\pi \sim \text{PHz}$) where the massive $Z^6$ prefactor provides exceptional sensitivity in the ultrahigh-frequency domain ($10^7\text{--}10^9\text{ Hz}$)~\cite{Ludlow:2015}. In contrast, Rydberg alkali ensembles (such as $^{87}\text{Rb}$ at $n \sim 45$) provide millimeter-wave transitions ($\omega_0/2\pi \sim 10\text{--}100\text{ GHz}$) and large dipole moments ($|\mathbf{d}| \sim 10^3\,e a_0$) well matched to acoustic and microwave gravitational-wave frequencies ($10^4\text{--}10^6\text{ Hz}$)~\cite{Gallagher:1994,2014PhRvD..90j2005G}. Although single-atom emission is suppressed at high $n$ by $n^{-8}$, large trapped ensembles ($N_{\text{at}} \sim 10^8$) coupled to high-finesse cavities ($F_P \sim 10^4$) compensate via collective atom-cavity emission.

\subsection{Noise Budget and Fundamental Detection Limits}

To establish the ultimate strain sensitivity of this setup, we have to weigh our subtle gravitational-wave signal against environmental noise and various decoherence pathways. For high-$n$ Rydberg states, the main culprit is ambient blackbody radiation (BBR), which drives thermal photo-excitation and ionization at a rate $W_{\text{BBR}} \propto T_{\text{env}} n^{-2}$~\cite{Gallagher:1994}. However, operating the cavity setup inside a standard dilution refrigerator at cryogenic temperatures ($T_{\text{env}} \approx 10\text{ mK}$) suppresses this BBR transition rate to $W_{\text{BBR}} \lesssim 10\text{ s}^{-1}$. This renders thermal decoherence completely negligible compared to the lightning-fast, cavity-enhanced Purcell decay. Similarly, ambient magnetic field noise can induce first-order Zeeman shifts ($\delta\omega_Z \approx \mu_B \delta B/\hbar$), but we can cleanly mitigate this by selecting magnetically insensitive clock transitions ($m_F = 0 \to m_F' = 0$). For the optical and UV transitions in HCIs, the energy gap is so huge ($\hbar\omega_0 \gg k_B T_{\text{env}}$) that thermal photon occupancy is naturally zero, even at room temperature.

With environmental decoherence safely out of the picture, the fundamental limit on our measurement is set purely by photon-counting shot noise at the detectors. For an ensemble of $N_{\text{at}}$ emitters, the total number of photons registered over a measurement duration $T_{\text{int}}$ is roughly $N_{\text{tot}} \sim \eta_{\text{eff}} N_{\text{at}} F_P \Gamma_0 T_{\text{int}}$. The overall efficiency parameter $\eta_{\text{eff}} \equiv \eta_{\text{geom}}\eta_{\text{det}} \le 1$ accounts for how well the cavity mode overlaps with the atoms, the mirror output coupling, and the finite quantum efficiency of the detectors. In state-of-the-art setups, this composite efficiency routinely reaches $0.60\text{--}0.75$ using superconducting nanowire single-photon detectors (SNSPDs) in the optical regime~\cite{Marsili:2013} and quantum-limited amplifiers in the microwave domain. 

Here, we must be careful regarding the integration time. Because our quadrupolar GW signature $\mathcal{A}_{\text{obs}}(t)$ oscillates at the frequency $\Omega$, simply averaging the differential photon counts over any duration longer than half a GW period ($\pi/\Omega$) would cause the positive and negative asymmetries to cancel out, destroying the signal. To prevent this, the differential detector output must be continuously monitored and phase-sensitively demodulated at the target frequency $\Omega$ (akin to lock-in amplification). Therefore, $T_{\text{int}}$ represents the effective integration time of this properly demodulated signal stream.

The Poisson statistical uncertainty in this normalized asymmetry observable is $\delta\mathcal{A} \approx 1/\sqrt{N_{\text{tot}}}$. Notice that because the detector efficiency $\eta_{\text{eff}}$ scales both the differential signal and the total photon count in the same way, it elegantly divides out of the normalized observable's mean. Setting our signal amplitude $C_{\mathbf{k}_c} \approx (\omega_0/\Omega) h$ equal to this shot-noise floor gives us the minimum detectable strain sensitivity at a unit signal-to-noise ratio ($\text{SNR}=1$):
\begin{equation}
h_{\min} \approx \left( \frac{\Omega}{\omega_0} \right) \frac{1}{\sqrt{\eta_{\text{eff}} N_{\text{at}} F_P \Gamma_0 T_{\text{int}}}}.
\label{eq:StrainSensitivity}
\end{equation}

Evaluating Eq.~\eqref{eq:StrainSensitivity} with the physical parameters summarized in Table~\ref{tab:exp_params} highlights the distinct advantages of our two operational regimes. For highly charged ions operating in the megahertz window, we project an exceptional strain sensitivity of $h_{\min} \sim 2.1 \times 10^{-17}\text{ Hz}^{-1/2}$. Meanwhile, for Rydberg alkali ensembles tailored to the $100\text{ kHz}$ acoustic band, we project a sensitivity of $h_{\min} \sim 3.8 \times 10^{-14}\text{ Hz}^{-1/2}$. This establishes differential cross-cavity spontaneous emission as a highly viable, background-resilient platform for probing the ultrahigh-frequency gravitational-wave frontier.

\section{Conclusion}

In this work, we have developed a comprehensive quantum electrodynamical framework to describe how an atom spontaneously emits light in a dynamical spacetime. By quantizing the electromagnetic field within a classical gravitational-wave background, we found that the rippling spacetime fundamentally dresses the vacuum modes, splitting them into a coherent Floquet ladder of sidebands separated by the GW frequency $\Omega$. While the total, angle-integrated spontaneous emission rate remains completely unchanged to first order in the strain $h$, the gravitational wave leaves a distinct physical footprint: it steers the emitted photons into a pronounced spin-2 quadrupolar pattern, $\Delta(\theta,\phi) \propto (\omega_0/\Omega) h \sin^2\theta \cos(2\phi)$. At finite interaction times, this spatial redistribution is accompanied by a characteristic red-blue spectral asymmetry.

To translate this theoretical vacuum anisotropy into a measurable laboratory signal, we proposed a cavity-QED detection protocol based on a pair of orthogonal, rigid microcavities. By leveraging the Purcell effect, this architecture effectively acts as a funnel, mapping the highly directional vacuum modulation directly into a time-dependent, differential photon count. Crucially, continuously monitoring and phase-demodulating this balanced readout cleanly cancels isotropic blackbody backgrounds and common-mode technical noise. 

When we evaluated the underlying atomic scaling laws, two distinct, highly optimized operating regimes emerged. For the acoustic and microwave bands ($10^4$--$10^6\text{ Hz}$), cryogenic Rydberg alkali ensembles offer massive dipole moments capable of reaching strain sensitivities of $h_{\min} \sim 3.8 \times 10^{-14}\text{ Hz}^{-1/2}$. For the higher-frequency megahertz-to-gigahertz regime ($10^7$--$10^9\text{ Hz}$), the severe $Z^6 scaling$ of highly charged ions provides exceptional projected sensitivities reaching $h_{\min} \sim 2.1 \times 10^{-17}\text{ Hz}^{-1/2}$.

Ultimately, these results demonstrate that precision cavity QED systems can serve as robust, highly sensitive detectors for gravitational waves, naturally filtering out environmental noise through their differential design. By bridging the gap between quantum optics and general relativity, this cross-cavity platform provides a promising new laboratory pathway to explore light-matter interactions in curved spacetime and to search for the elusive ultrahigh-frequency gravitational waves predicted by early-Universe cosmology.

\section*{Acknowledgments}
Author used Gemini Pro to improve the text of paper. 

\bibliographystyle{DU-PRSTY}   
\bibliography{gw_rydberg_refs}

\begin{thebibliography}{10}

\bibitem{LIGOScientific:2016aoc}
{Abbott}, R., \& {\textit{et al.}} 2016, Phys. Rev. Lett., {116},  061102.

\bibitem{2023PhRvX..13a1048A}
{Abbott}, R., \& {\textit{et al.}} 2023, Physical Review X, {13},  011048.

\bibitem{2013PhRvL.110g1105A}
{Arvanitaki}, A., \& {Geraci}, A.~A. 2013, Physical Review Lett., {110},
  071105.

\bibitem{2025LRR....28...10A}
{Aggarwal}, N., {et~al.} 2025, Living Reviews in Relativity, {28},  10.

\bibitem{Gehrman2023}
Gehrman, T.~C., Haghi, B. S.~E., Sinha, K., \& Xu, T. 2023, Journal of
  Cosmology and Astroparticle Physics, {2023},  001.

\bibitem{Brito2020}
Brito, R., Cardoso, V., \& Pani, P. 2020, {Superradiance: New Frontiers in
  Black Hole Physics}, Vol.~906 of {Lecture Notes in Physics},  Springer.

\bibitem{2014PhRvD..90j2005G}
{Goryachev}, M., \& {Tobar}, M.~E. 2014, Physical Review D, {90},  102005.

\bibitem{PhysRevD.108.L061303}
Bringmann, T., Domcke, V., Fuchs, E., \& Kopp, J. 2023, Phys. Rev. D, {108},
  L061303.

\bibitem{PhysRevLett.132.190001}
Ye, J., \& Zoller, P. 2024, Phys. Rev. Lett., {132},  190001.

\bibitem{Ludlow:2015}
Ludlow, A.~D., {et~al.} 2015, Rev. Mod. Phys., {87},  37.

\bibitem{Bothwell:2022}
Bothwell, T., {et~al.} 2022, Nature, {602},  420.

\bibitem{Graham:2013gfa}
{Graham}, P.~W., {Hogan}, J.~M., {Kasevich}, M.~A., \& {Rajendran}, S. 2013,
  Physical Review Lett., {110},  171102.

\bibitem{Baynham2025}
Baynham, C. F.~A., {et~al.} 2025, arXiv preprint arXiv:2504.09158,

\bibitem{Sedlacek2012}
Sedlacek, J.~A., {et~al.} 2012, Nature Physics, {8},  819.

\bibitem{Paczos2026GWImprints}
Paczos, J., {et~al.} 2026, Physical Review Lett., {136},  113201.

\bibitem{PhysRevD.105.024053}
Chen, B.-H., \& Chiou, D.-W. 2022, Phys. Rev. D, {105},  024053.

\bibitem{Prokopec_2023}
Prokopec, T. 2023, Classical and Quantum Gravity, {40},  035007.

\bibitem{Birrell1982}
Birrell, N.~D., \& Davies, P. C.~W. 1982, {Quantum Fields in Curved Space},
  Cambridge University Press.

\bibitem{Wald:1995yp}
Wald, R.~M. 1995, {{Quantum Field Theory in Curved Space-Time and Black Hole
  Thermodynamics}}, {Chicago Lectures in Physics},  University of Chicago
  Press.

\bibitem{Purcell1946}
Purcell, E.~M. 1946, Physical Review, {69},  681.

\bibitem{Haroche2006}
Haroche, S., \& Raimond, J.-M. 2006, {Exploring the Quantum: Atoms, Cavities,
  and Photons},  Oxford University Press.

\bibitem{ScullyZubairy1997}
Scully, M.~O., \& Zubairy, M.~S. 1997, {Quantum Optics},  Cambridge University
  Press.

\bibitem{Sakurai2020}
Sakurai, J.~J., \& Napolitano, J. 2020, {Modern Quantum Mechanics}, , 3rd ed.
  Cambridge University Press.

\bibitem{Milonni1994}
Milonni, P.~W. 1994, {The Quantum Vacuum: An Introduction to Quantum
  Electrodynamics},  Academic Press.

\bibitem{Gallagher:1994}
Gallagher, T.~F. 1994, {Rydberg Atoms},  Cambridge University Press.

\bibitem{Marsili:2013}
Marsili, F., {et~al.} 2013, Nature Photonics, {7},  210.

\end{thebibliography}

\end{document}